\documentclass[pdflatex,iicol,sn-mathphys-num]{sn-jnl}

\usepackage[T1]{fontenc}
\usepackage{textcomp}
\usepackage{wrapfig}
\usepackage{subcaption}
\usepackage{inconsolata}
\usepackage{xspace}
\usepackage{verbatim}
\usepackage{fancyvrb}
\usepackage{fvextra}
\usepackage{graphicx}%
\usepackage[labelfont=bf]{caption}
\usepackage{multirow}%
\usepackage{amsmath,amssymb,amsfonts}%
\usepackage{amsthm}%
\usepackage{mathrsfs}%
\usepackage[title]{appendix}%
\usepackage{xcolor}%
\usepackage{textcomp}%
\usepackage{manyfoot}%
\usepackage{booktabs}%
\usepackage{algorithm}%
\usepackage{algorithmicx}%
\usepackage{algpseudocode}%
\usepackage{listings}%
\usepackage{csquotes}
\usepackage{comment}
\usepackage{enumerate}
\usepackage{hyperref,cleveref}
\usepackage{float}
\usepackage{placeins}
\usepackage{mdframed}
\usepackage{tcolorbox}
\usepackage{flushend}
\usepackage{tikz}
\usetikzlibrary{positioning, arrows.meta}

\definecolor{backcolour}{rgb}{0.9,0.9,0.9}

\lstdefinestyle{mystyle}{
    backgroundcolor=\color{backcolour}
}

\newcommand{\eg}{e.g.,\xspace}

\newcommand{\ie}{i.e.,\xspace}
\newcommand{\viz}{viz.,\xspace}
\newcommand{\cd}[1]{\texttt{\small #1}\xspace} 
\newcommand{\wrt}{w.r.t.\xspace}

\newcommand{\A}{\texttt{A}\xspace}
\newcommand{\I}{\texttt{I}\xspace}
\newcommand{\R}{\texttt{R}\xspace}
\newcommand{\RP}{\texttt{R+}\xspace}
\newcommand{\rp}{\texttt{r+}\xspace}
\newcommand{\D}{\texttt{D}\xspace}
\newcommand{\E}{\texttt{E}\xspace}
\newcommand{\HF}{\texttt{H}\xspace}
\newcommand{\eclypse}{\textit{ECLYPSE}\xspace}
\newcommand{\mcs}{Multi-Criteria Solver\xspace}
\newcommand{\ee}{Energy Enhancer\xspace}
\newcommand{\fe}{Failure Enhancer\xspace}

\newcommand{\mycomment}[3]{\begin{mdframed}[backgroundcolor=#3]\textbf{#1:} #2\end{mdframed}}
\newcommand{\tbd}[1]{\mycomment{TBD}{#1}{red!20}}

\DefineVerbatimEnvironment{code}{Verbatim}
{samepage=true, fontfamily=zi4, fontsize=\fontsize{8.2}{10}\selectfont, frame=single, framesep=1mm, framerule=0.1pt, rulecolor=\color{gray}, commandchars=\\\{\}, vspace=3pt}
\DefineVerbatimEnvironment{codenum}{Verbatim}
{fontfamily=zi4, numbers=left, numbersep=5pt, numberblanklines=false, firstnumber=last, tabsize=2, fontsize=\fontsize{8}{10}\selectfont, frame=single, framesep=1mm, framerule=0.1pt, rulecolor=\color{gray}, gobble=4, vspace=3pt}

\theoremstyle{thmstyleone}%
\theoremstyle{thmstyletwo}%

\theoremstyle{thmstylethree}%

\begin{document}

\title[FREEDA]{Failure-Resilient and Carbon-Efficient Deployment of Microservices over the Cloud-Edge Continuum}


\author[1]{\fnm{Francisco} \sur{Ponce}}

\author[2]{\fnm{Simone} \sur{Gazza}}

\author[3]{\fnm{Andrea} \sur{D'Iapico}}

\author[2]{\fnm{Roberto} \sur{Amadini}}
\author[1]{\fnm{Antonio} \sur{Brogi}}
\author[1]{\fnm{Stefano} \sur{Forti}}
\author[2]{\fnm{Saverio} \sur{Giallorenzo}}
\author[3]{\fnm{Pierluigi} \sur{Plebani}}
\author[1]{\fnm{Davide} \sur{Usai}}
\author[3]{\fnm{Monica} \sur{Vitali}}
\author[2]{\fnm{Gianluigi} \sur{Zavattaro}}
\author*[1]{\fnm{Jacopo}
\sur{Soldani}}\email{jacopo.soldani@unipi.it}

\affil[1]{\orgname{University of Pisa}, \orgaddress{\city{Pisa}, \country{Italy}}}
\affil[2]{\orgname{University of Bologna}, \orgaddress{\city{Bologna}, \country{Italy}}}
\affil[3]{\orgname{Politecnico di Milano}, \orgaddress{\city{Milano}, \country{Italy}}}


\abstract{Deploying microservice-based applications (MSAs) on heterogeneous and dynamic Cloud-Edge infrastructures requires balancing conflicting objectives, such as failure resilience, performance, and environmental sustainability. In this article, we introduce the FREEDA toolchain, designed to automate the failure-resilient and carbon-efficient deployment of MSAs over the Cloud-Edge Continuum.
The FREEDA toolchain continuously adapts deployment configurations to changing operational conditions, resource availability, and sustainability constraints, aiming to maintain the MSA quality and service continuity while reducing carbon emissions. We also introduce an experimental suite using diverse simulated and emulated scenarios to validate the effectiveness of the toolchain against real-world challenges, including resource exhaustion, node failures, and carbon intensity fluctuations. The results demonstrate FREEDA’s capability to autonomously reconfigure deployments by migrating services, adjusting flavour selections, or rebalancing workloads, successfully achieving an optimal balance among resilience, efficiency, and environmental impact.
}

\keywords{Cloud-Edge Continuum, Microservices, Failure Resilience, Environmental Sustainability}



\maketitle

\section{Introduction}
\label{sec:introduction}


The rapidly expanding capabilities of smart, connected IoT devices necessitate an evolution of cloud computing into large-scale, pervasive, and distributed environments. These environments must minimize unnecessary latencies while fully leveraging the computing resources available at the network edge \cite{Ferrer2019}. The infrastructure enabling this Cloud-Edge continuum will inevitably be highly heterogeneous and dynamic.
Heterogeneity arises from the diversity of devices involved, which feature varying levels of compute and storage capacity, rely on different deployment and software technologies, and communicate through multiple protocols. Variability, on the other hand, comes from both dynamism (\eg nodes joining or leaving the infrastructure, fluctuating workloads) and uncertainty (\eg unstable end-to-end connectivity or hardware failures). Together, this heterogeneity and variability amplify the challenges of preventing quality-of-service (QoS) degradation and handling faults \cite{Gaglianese2023}.

Simultaneously, the widespread adoption of microservices in delivering enterprise solutions has increased the need for effective deployment strategies across the Cloud-Edge continuum. MSAs consist of multiple interdependent services, with varying deployment requirements. These may include cost constraints for Cloud-Edge resource rental, specific hardware and software dependencies, security considerations, and strict QoS demands such as low latency, sufficient bandwidth, and high availability.
Given the inherent complexity of both MSAs and Cloud-Edge infrastructures, as well as the volatility of edge nodes and services, failures must be treated as first-class concerns. Not only must engineers account for individual service or node failures, but also for cascading failures, where the malfunction of one component propagates to others. To ensure that deployed MSAs consistently meet their QoS objectives, DevOps engineers must proactively design with these risks in mind, embedding resilience into deployment and management strategies \cite{Soldani2022_AnomalyDetection}.

The need for the deployment of resilient applications is increasingly aligned with growing global interest in sustainability and environmental responsibility. Initiatives such as the EU strategy of ``building a climate-neutral, green, fair, and social Europe'' \cite{EU2019}, which promotes the environmentally sustainable growth of European industries, including IT \cite{EU2022}, reflect this broader trend. Consequently, environmental sustainability is becoming a key consideration when deploying applications over Cloud-Edge infrastructures, \eg by seeking to minimize the carbon footprint of deployed MSAs.

However, sustainability objectives may conflict with other deployment requirements, such as failure resilience. For example, carbon emissions can be reduced by consolidating services, \ie deploying a single instance of each service on nodes located close to one another, or even on the same node. Yet, this approach undermines resilience to failures. Conversely, replicating services across multiple, geographically distributed nodes strengthens resilience but increases energy consumption and carbon emissions.
This trade-off highlights the complexity of deploying MSAs in large-scale, heterogeneous, and dynamic Cloud-Edge environments. Effective deployment support must therefore balance multiple, and often conflicting, requirements, including sustainability, resilience, cost, security, and QoS.
Moreover, the variability of Cloud-Edge infrastructure over time (\eg nodes joining, leaving, failing, or becoming overloaded) must also be accounted for. This is particularly critical for already deployed MSAs, which are frequently updated and continuously delivered through CI/CD pipelines.

Several efforts have explored application placement strategies to reduce energy consumption or carbon emissions \cite{abbasikhazaei2022,Ahvar2021,Mohammad2021}. However, only a few have addressed adaptive deployment, \ie dynamically adjusting application components and their placement in response to changing contexts, objectives, or workloads. For example, Forti and Brogi \cite{Forti2021} propose deploying application components in different functionally equivalent \textit{flavours} according to operator preferences and cost objectives, employing a greedy strategy to minimize operational expenses. Yet, their approach does not incorporate sustainability or carbon awareness. Other works \cite{Vitali2022,Vitali2023-Enriching} exploit flavours in the context of MSAs to adapt workflows and mitigate environmental impact, but they do not address deployment challenges in heterogeneous, distributed infrastructures.

To this end, the FREEDA research project \cite{FREEDA-Frame2024} aims at supporting DevOps teams managing highly heterogeneous and dynamic environments. The main goal is indeed to address the failure-resilient, energy-aware, and explainable deployment of microservice-based Applications over Cloud-Edge infrastructures. FREEDA automates the optimal selection of component flavours and their feasible placement across the Cloud-Edge continuum, taking into account dependencies, topology, resource availability, costs, and sustainability constraints. Its primary objective is to determine deployment configurations that satisfy application requirements while ensuring both carbon efficiency and resilience to failures.

The baseline idea is that supporting component flavours in Cloud-Edge deployments is essential to achieving high levels of automation and flexibility. Flavours represent functionally equivalent implementations of a component that differ in performance, resource requirements, or carbon footprint. Leveraging these alternatives allows operators to optimize application performance while satisfying specific operational constraints, such as carbon budgets, cost, or resilience requirements.

In this article, we present the FREEDA toolchain, which automates the optimal selection of component flavours and their feasible placement, continuously adapting deployment configurations to changing operational conditions, resource availability, and sustainability constraints. The FREEDA toolchain also manages the reconfigurations needed to address failures or to improve overall carbon efficiency. Indeed, when presented with a plethora of equivalent adaptations, the more components that need changing the higher becomes the overhead imposed by the deployment adaptation (service downtime, performance degradation, possible instability). To this end, we improve on the state of the art~\cite{toit} by revising the existing FREEDA model, underlying the FREEDA toolchain, introducing new constraints for the minimization of deployment changes (equivalently to maximizing the parts of the system that remain unchanged).

We also present an experimental suite to validate the effectiveness of the FREEDA toolchain. This setup uses diverse, controlled simulated and emulated scenarios to test FREEDA against real-world challenges, including resource exhaustion, node failures, carbon intensity fluctuations, and multi-objective trade-offs between performance and sustainability. The results demonstrate FREEDA's capability to autonomously reconfigure deployments---by migrating services, adjusting flavour selections, or rebalancing workloads---to achieve an optimal balance among resilience, efficiency, and environmental impact.

\smallskip \noindent
The remainder of this article is structured as follows.
\Cref{sec:background} retakes and extends our constraint-based deployment model.
\Cref{sec:toolchain} provides an overview of the FREEDA toolchain.
\Cref{sec:simulation} presents and discusses the simulation results, while \Cref{sec:emulation} focuses on the emulation results.
Finally, \Cref{sec:related-work,sec:conclusion} review related work and provide concluding remarks, respectively.

This paper extends our previous works \cite{toit,FREEDA_Failure_Resilience2025}, where we first presented our constraint model \cite{toit} and the ideas behind FREEDA's failure enhancer \cite{FREEDA_Failure_Resilience2025}. This article extends the deployment model (\Cref{sec:background}) and provides a thorough description of the full FREEDA toolchain (\Cref{sec:toolchain}). Additionally, this article provides brand-new experiments (\Cref{sec:simulation,sec:emulation}) to assess how effectively FREEDA can support the sustainable and failure-resilient deployment of MSAs over Cloud-Edge infrastructures.

\section{Deployment Model} 
\label{sec:background}

\paragraph{The FREEDA Model}
MSAs decompose applications into multiple software components. However, besides
having to deal with the deployment of components, adaptability under different
deployment contexts compels the availability of multiple versions---or
\emph{flavours}~\cite{toit, lopstr2024}---of the components, each with specific
functional and non-functional properties. Thus, deploying such applications
entails two interdependent decisions: (i) select an appropriate flavour for
each component, and (ii) map those components to nodes across a heterogeneous
infrastructure ranging from powerful Cloud to constrained Edge/IoT nodes.

The deployment must simultaneously account for a variety of requirements and objectives, including QoS constraints such as latency, bandwidth, and availability; operational concerns such as budget and cost efficiency; and increasingly pressing environmental considerations, such as energy consumption and carbon emissions. In addition, functional dependencies among components and infrastructural limitations must be respected. Exploring this vast, multi-dimensional solution space is combinatorially complex and often infeasible without systematic, automated support, particularly given the conflicts that may arise among competing objectives.

To address this challenge, in previous work~\cite{toit}, we introduced a constraint optimisation model that provides the mathematical foundation of the FREEDA framework. The goal of this model is to transform high-level deployment specifications---concerning application components, their flavours, deployment requirements, and infrastructure characteristics---into a rigorous optimisation problem whose solutions correspond to feasible deployments. Specifically, the model jointly: (i) selects each component and its respective flavour and assigns it to a node, (ii) ensures compliance with all deployment requirements expressed in the FREEDA YAML specification~\cite{freedayamlspec}, including energy and carbon budgets, and (iii) prioritises the deployment of the most powerful flavours whenever possible, in line with application owners' preferences.
The full formal specification of the model~\cite[Section 4]{toit} is articulated into four essential parts---parameters, variables, constraints, and the objective function---which are described informally in the following paragraphs. 

Parameters provide the input data for each deployment instance, capturing the set of application components and their associated flavours, dependencies between components, resource requirements (both consumable, such as CPU, RAM, and storage, and non-consumable, such as availability or security), infrastructure specifications (\eg node capacities, link latency and availability, and per-resource monetary and carbon costs), and budget thresholds for expenditure and emissions. Each flavour has an \emph{importance} value: the higher the value, the more powerful the flavour.

Variables encode the deployment decisions as binary indicators $D^c_{i, j} \in
\{0,1\}$ where $c$, $i$, and $j$ respectively denote a component, a flavour (of
$c$), a node in the infrastructure. The constraint solver sets $D^c_{i, j} = 1$
if and only if $c$ is deployed in flavour $i$ on $j$. This representation is
solver-agnostic; in fact, various solving technologies can execute the model,
including constraint programming (CP), mixed-integer linear programming (MIP),
or Boolean Satisfiability (SAT).

Constraints formalise the validity conditions of a deployment. They guarantee, for example, that each component is deployed at most once (in one flavour on one node), that mandatory components are always deployed, that required dependencies are satisfied with flavours of sufficient power, and that non-essential components are never deployed in isolation. Resource constraints ensure that aggregate consumptions do not exceed node capacities, while network link constraints ensure latency and availability requirements are satisfied. Budget constraints enforce compliance with financial and carbon limitations. 

The objective function maximizes the importance of deployed flavours rather than directly minimizing costs or emissions, thereby avoiding ``empty deployments'' where no component is deployed to reduce expenses. Budgets are enforced as hard constraints, while the optimization prioritizes higher-importance flavours wherever feasible. The user can manually define a flavour's importance or select it from a set of built-in importance policies.

\paragraph{Minimizing Deployment Changes}
In this work, we advance the state of the art~\cite{toit} by focussing on  
reducing \emph{changes} across successive deployments. Indeed, while one needs to adapt an MSAs to changing conditions (like monetary and energy budgets and software and infrastructure failures), minimising the number of changes across successive deployments is critical for maintaining overall system availability (\eg downtime), performance, and stability. This additional objective introduces a new minimisation criterion for the solver. Specifically, when a new deployment is requested, the solver is instructed to minimise the number of components that either switch flavour or are relocated to a different node \wrt the current deployment. We capture this property by tracking each component $c$ deployed in the current configuration (\ie those for which $D^c_{i, j}=1$ for some flavour $i$ and node $j$), aiming to preserve their deployment in the subsequent configuration whenever possible. In other words, if $D^c_{i, j}=1$ in the current deployment, the solver attempts to enforce $D^c_{i, j}=1$ in the new deployment as well.

Formally, this requirement corresponds to maximizing the number of components that remain deployed with the same flavour on the same node, that is:
\begin{equation}\label{eq:newObjFun}
    \max \sum_{\bar{D}^c_{i, j} = 1}  D^c_{i, j}
\end{equation}
where $\bar{D}$ denotes the already computed matrix of decision variable \emph{values} from the current deployment, and $D$ represents the matrix for the new deployment. Flavour changes and node relocations are treated equivalently because both operations ultimately require redeploying the component. 
\tbd{commento su caso degenere nel re-deployment (qui o nella discussion)}

\section{FREEDA Toolchain}
\label{sec:toolchain}
\begin{figure*}[t]
    \centering
    \includegraphics[trim=6cm 0cm 6cm 0cm, width=.95\textwidth]{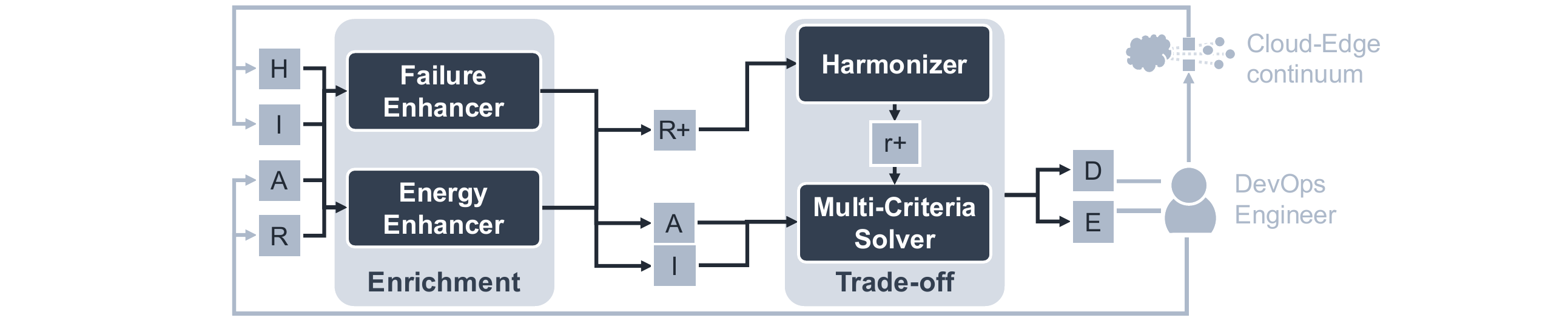}
    \caption{A bird's-eye view of FREEDA's toolchain}
    \label{fig:freeda_approach}
\end{figure*}

FREEDA aims to address the demand for DevOps support in deploying an MSA \A over a Cloud-Edge infrastructure \I, guided by a set of deployment requirements \R, as it can be seen in \Cref{fig:freeda_approach}. The infrastructure description \I includes information such as resource utilization costs, hardware and software capabilities of nodes, and their current load and availability. In contrast, the requirements \R specify the needs of the services composing \A, including hardware and software dependencies, network QoS, security, and failure resilience. Additionally, \R may define deployment budgets, both monetary (\ie the maximum affordable cost) and sustainability-related (\ie the maximum permissible energy consumption and carbon emissions).

The FREEDA toolchain generates a deployment plan \D for \A over \I by holistically identifying a trade-off among the requirements in \R. As shown in \Cref{fig:freeda_approach}, the proposed solution is divided into two main phases (\textit{Enrichment} and \textit{Trade-Off}) each supported by two main components. The first phase enhances the failure resilience and environmental sustainability of the application by leveraging historical data \HF, which includes information from current and past deployments of \A (\eg logs) as well as monitoring data from \I (\eg node availability or load). This step produces an enriched set of requirements \RP that includes carbon-aware and failure-resiliency considerations. This automated enrichment reduce the design effort required from DevOps Engineers by ensuring the satisfaction of these non-functional aspects in the deployment of the application without the need for direct intervention.

In the second phase, the trade-off analysis tools process \A and \RP to produce a deployment plan \D together with an explanation \E. The explanation clarifies why \D represents the best trade-off among the multiple---and potentially conflicting---deployment requirements. Importantly, \E also documents the reasoning behind the enrichment process, detailing how and why \RP was transformed into \rp. This ensures that DevOps engineers are not only informed of the final deployment plan but also understand the rationale for any adjustments made to the application specification and its requirements.

Using these inputs, FREEDA will provide the DevOps Engineer with a valid deployment able to fulfill all the expressed constraints. The core of the methodology is the \mcs component. In this context, the explanation \E corresponds to the constraint model itself, which, when executed by the \mcs, produces the optimal deployment plan \D.

Each component of the FREEDA toolchain illustrated in \Cref{fig:freeda_approach} is described hereafter.

\subsection{Failure Enhancer}
\label{subsec:failure-enhancer}
The Failure Enhancer is responsible for generating soft deployment requirements within \R that aim to improve the resilience of microservices prior to deployment. These requirements help, for instance, to avoid deploying services on nodes known to fail under certain load conditions or on nodes whose failure is predicted to occur soon based on available historical data.
As described in \cite{FREEDA_Failure_Resilience2025}, the Failure Enhancer produces three types of constraints: \textit{Affinity}, \textit{Anti-affinity}, and \textit{Avoid}.
An \textit{Affinity} constraint suggests deploying components \cd{C} and \cd{S} in their current flavour closer, \eg on the same node. 
An \textit{Anti-affinity} constraint suggests avoiding placing components \cd{C} and \cd{S} in their current flavour onto the same node.
While an \textit{Avoid} constraint suggests avoiding placing component \cd{C} in its current flavour \cd{FC} onto node \cd{N}.
\Cref{fig:FE_clauses} showcases the Prolog clauses used to generate the above-described soft constraints, which are discussed hereafter.

\begin{figure*}[ht]
    \centering
\begin{codenum}[firstnumber=1]
    suggested(affinity(d(C,FC),d(S,FS))) :-
        deployedTo(C,FC,N), deployedTo(S,FS,M), dif(C,S), dif(N,M), 
        timeoutEvent(C,S,T),
        \+( congested(N,M,T); disconnected(N,T); disconnected(M,T) ).
        
    suggested(avoid(d(C,FC),N)) :-
        deployedTo(C,FC,N), deployedTo(S,_,M), dif(C,S), dif(N,M),
        timeoutEvent(C,S,T),
        ( congested(N,M,T); disconnected(N,T) ).

    suggested(antiaffinity(d(C,FC),d(S,FS))) :-
        deployedTo(C,FC,N), 
        ( unreachable(C,T); internal(C,T) ),
        overloaded(N,R,T), race(N,R,C,FC,S,FS,T).
    
    suggested(avoid(d(C,FC),N)) :-
        deployedTo(C,FC,N),
        ( unreachable(C,T); internal(C,T) ),
        ( (overloaded(N,_,T), \+ race(N,_,C,FC,_,_,T)) ; disconnected(N,T) ).
\end{codenum}
    \caption{Failure Enhancer Prolog clauses presented in \cite{FREEDA_Failure_Resilience2025}}
    \label{fig:FE_clauses}
\end{figure*}

\paragraph{Knowledge representation}
The current MSA deployment information is denoted via Prolog facts like \cd{deployedTo(C,F,N)}, indicating that component \cd{C} is deployed in its flavour \cd{F} to node \cd{N}. 
%
From an MSA failure perspective, the Failure Enhancer assumes that timeout events between components \cd{C1} and \cd{C2} are denoted via timestamped facts like \cd{timeoutEvent(C1,C2,Timestamp)}. Besides, internal error and unreachability for a component \cd{C} are denoted via facts like \cd{internal(C,Timestamp)} and \cd{unreachable(C,Timestamp)}, respectively. 

On the other hand, considering infrastructure logging, the predicate \cd{congested(N,M,T)} identifies that link congestion between nodes \cd{N} and \cd{M} occurred at time \cd{T}. Likewise, the predicate \cd{disconnected(N,T)} holds true if node \cd{N} incurred a network disconnection at time \cd{T}.
Predicate \cd{overloaded(N,R,T)} denotes the situation in which a specific resource \cd{R} (\eg RAM, CPU, HDD) was subject to overloading at time \cd{T}.
Last, predicate \cd{race(N,R,C,FC,S,FS,T)} denotes a situation at time \cd{T} in which components \cd{C} (flavoured \cd{FC}) and \cd{S} (flavoured \cd{FS}) were racing for resource \cd{R} on the same node \cd{N}.

\paragraph{Enhancing failure resilience}
Based on the above, the Failure Enhancer generates a set of suggested soft constraints to improve the resilience of the current deployment by embedding rules of thumb to improve placement decisions.
Indeed, it finds all distinct suggested constraints by checking which clauses of predicate \cd{suggested/1} fire in the considered knowledge base. We here discuss the four clauses of such a predicate shown in \Cref{fig:FE_clauses}, noting, however, that they can be easily extended or refined to account for more MSA failures and/or network conditions.

The first clause of \cd{suggested/1} (lines 1--4) identifies that an interaction between \cd{dif}ferent components \cd{C} (flavoured \cd{FC}) and \cd{S} (flavoured \cd{FS}), deployed to two distinct nodes \cd{N} and \cd{M}, respectively (line 2), went through a timeout event (line 3), despite no congestion or disconnection events occurred (line 4). 
To avoid this from happening again, the Failure Enhancer suggests deploying \cd{C} and \cd{S} closer, by adding an affinity constraint between the two components in their current flavours, \viz \cd{affinity(d(C,FC),d(S,SF))}.

The second clause of \cd{suggested/1} (lines 5--8) identifies that a timeout event at time \cd{T} at component \cd{C} in its flavour \cd{FC} deployed to node \cd{N} and involving component \cd{S} deployed to a distinct node \cd{M} (line 6), might have been caused by network congestion between \cd{N} and \cd{M} or by disconnection of node \cd{N} (line 8). The Failure Enhancer suggests avoiding placing \cd{C} (flavoured \cd{FC}) onto node \cd{N} by including a node avoidance constraint, \viz \cd{avoid(d(C,FC),N)} (line 5). Indeed, the link between nodes \cd{N} and \cd{M} might be continuously subject to congestion or faulty. A symmetric clause (not shown) exists to handle the symmetric situation (\ie congestion or disconnection of node \cd{M}) by suggesting an \cd{avoid(d(S,FS),M)} constraint.

The third clause of \cd{suggested/1} (lines 9--12) identifies that component \cd{C} (flavoured \cd{FC}), deployed to node \cd{N} (line 10) was either unreachable or experiencing an internal error at time \cd{T} (line 11). This failure overlapped with node \cd{N} overloading of resource \cd{R}, due to another component \cd{S} (flavoured \cd{FS}) racing with \cd{C} for the resource \cd{R} (line 12).
The Failure Enhancer suggests avoiding placing \cd{C} and \cd{S} onto the same node through an \textit{Anti-affinity} constraint between the two components in their current flavours, \viz \cd{antiaffinity(d(C,FC),d(S,SF))} (line 9). 

The fourth and last clause of \cd{suggested/1} (lines 13--16) identifies component \cd{C} deployed to \cd{N} in its flavour \cd{FC} (line 14) went through a failure event (line 15), which was possibly due to node overloading (in absence of a race) or disconnection  (line 16). The Failure Enhancer suggests avoiding placing \cd{C} (flavoured \cd{FC}) onto node \cd{N} by including a node avoidance constraint, \viz \cd{avoid(d(C,FC),N)} (line 13). 

\subsection{\ee}
\label{subsec:energy-enhancer}

In coordination with the Failure Enhancer, the \ee generates soft deployment requirements within \R that aim to reduce the environmental impact of the MSA execution.

The \ee is a key component of the proposed solution, responsible for enabling carbon-efficient and environmentally conscious application deployments across the cloud continuum. This component ingests multiple inputs, including the \textit{Application Description} (i.e., the services to be deployed), the \textit{Infrastructure Description} (i.e., the available computing nodes), the current \textit{Grid Carbon Intensity}, and relevant \textit{Monitoring Metrics} that capture the runtime behavior of applications. Based on these inputs, the component produces a set of constraints that inform and guide the \mcs during the generation of a suitable deployment plan. Moreover, the \ee continuously improves its outputs by iteratively learning from past deployments, enabling the system to adapt to changes in both application behavior and infrastructure conditions.

The solution incorporates several key functionalities that work together to generate and refine environmentally aware deployment constraints. First, information about the carbon intensity of the underlying infrastructure is continuously gathered and processed. Instead of relying on instantaneous values, the system considers aggregated measurements over a recent observation window, resulting in more stable and representative data for decision-making. 

In parallel, the carbon footprint associated with application services and their inter-service communications is estimated by analysing historical monitoring data. This allows the system to enrich the initial deployment descriptions with energy profiles that capture both computational and communication-related energy consumption.

Using these enriched inputs, the solution derives deployment constraints that reflect the current application behavior and infrastructure conditions. These constraints are defined according to pre-established templates and rules, which can be extended to accommodate new types of green-aware policies as needed. By doing so, the system remains flexible and adaptable to evolving sustainability objectives.

To ensure that the generated constraints build upon prior knowledge, previously learned information is retrieved, refined, and incrementally updated over time. This continuous learning process helps maintain consistency across multiple deployment cycles, avoiding the loss of useful insights from past configurations.

Given the potentially large number of constraints that may emerge, the solution applies ranking and filtering techniques to retain only those with the highest expected impact on energy efficiency and carbon emissions. This prioritization step ensures that the resulting set of constraints remains both relevant and manageable.


\paragraph{Knowledge Base}
To make deployment decisions more effective, the generation of constraints should not rely solely on the most recent monitoring data but should also build on knowledge accumulated over time. For this purpose, the solution maintains a dedicated knowledge base that stores different types of information related to applications, their communications, the underlying infrastructure, and previously generated constraints.

The knowledge base keeps historical records of how each application service behaves in terms of energy consumption. Since a service may exhibit different energy profiles depending on where and how it is deployed, the system maintains information about typical ranges of its environmental footprint, including minimum, maximum, and average values observed from monitoring data collected during past deployments. In addition, the system stores information about the carbon impact of data exchanges between services. By analysing historical communication patterns, it builds a profile that reflects the environmental cost of interactions between different services across various deployments. The knowledge base also includes historical data on the carbon intensity of infrastructure nodes. Because the environmental characteristics of nodes can change over time, these records provide typical emission levels, which can be used to make better-informed placement decisions during future deployments.

Beyond raw monitoring data, the knowledge base preserves previously generated constraints. Each stored constraint contains information about its estimated environmental impact at the time of generation, along with a memory weight that reflects its current relevance. This weight decreases if the constraint is not regenerated over multiple iterations, ensuring that outdated information gradually loses influence.

The knowledge base is continuously enriched with newly collected data and recently generated constraints, while simultaneously updating the relevance of older ones. At each iteration, new deployment constraints are produced by observing the current data available, and valid past constraints are retrieved to complement them. This combined use of fresh observations and accumulated knowledge allows the system to generate more accurate, context-aware, and sustainable deployment decisions over time.

\paragraph{Enhancing environmental sustainability}
The \ee generates two types of constraints: \textit{Affinity} and \textit{Avoid}, with the same meaning of the constraints generated by the Failure Enhancer but with different conditions.

\textit{Avoid} constraints are related to the energy consumption of individual services. Their goal is limiting deployments that would lead to high energy usage or emissions. When a service, in a specific configuration, is known to consume excessive energy on a given node, a recommendation is generated to avoid that particular deployment.
\textit{Affinity} constraints target the energy overhead caused by service interactions. If two services exchange large amounts of data, deploying them on separate nodes may result in significant communication energy costs. In such cases, the system recommends co-locating the services to reduce this overhead.

The underlying logic for these two types of constraints can be expressed using Prolog clauses:

\begin{codenum}[firstnumber=1]
    suggested(avoid(d(C,FC), N)) :-
        highConsumptionService(C, FC, N).
    
    suggested(affinity(d(C,FC), d(S,FS))) :-
        dif(C, S),
        highConsumptionConnection(C, FC, S, FS).
\end{codenum}

The first clause produces recommendations to avoid placing certain service-flavour combinations on specific nodes when monitoring data indicates that such deployments are energy-inefficient. The generation of \textit{Avoid} constraints relies on historical information saved in the knowledge base. For each possible combination of service, flavour, and node, the system evaluates whether deploying that specific configuration would result in excessive energy use or carbon emissions. This is determined by combining the service's energy profile with the node's carbon intensity and comparing the result against a predefined threshold. Whenever the estimated impact exceeds the threshold, the system generates an \textit{Avoid} constraint to recommend avoiding the deployment of that service-flavour pair on the corresponding node. This ensures that environmentally costly placements are identified and excluded from the deployment planning process.

The second rule generates recommendations to place two interacting services on the same node when their communication pattern would otherwise lead to a high carbon footprint. The generation of \textit{Affinity} constraints is based on historical information saved in the knowledge base. For each potential combination of source service, its flavour, and destination service, the system evaluates whether their interaction is associated with high energy consumption. This assessment is performed by analyzing the energy profile of their communication and comparing it against a predefined threshold. If the estimated communication cost exceeds this value, the system generates an \textit{Affinity} constraint. These constraints guide the scheduler to place the involved services on the same node.
Together, these two constraints form the foundation for greener deployment.

\subsection{Harmonizer}
\label{subsec:harmonizer}

The Harmonizer component takes as input the enriched description of the application requirements, \RP. Its primary function is to identify and resolve potential conflicts among the soft requirements, guided by the priorities defined by DevOps engineers, whether emphasizing resilience, sustainability, or balance between them. After processing, the Harmonizer produces \rp, a refined subset (or, in some cases, the complete set) of requirements, which is subsequently forwarded to the \mcs for consideration in the next deployment phase.

\Cref{fig:harmonizer_example} illustrates a case of conflicting soft constraints generated by the Failure Enhancer and the Energy Enhancer. In this example, the Failure Enhancer suggests deploying the \textit{Frontend (Large flavour)} and \textit{Load Balancer (Large flavour)} services on different nodes to improve fault tolerance through \textit{antiaffinity}. Conversely, the Energy Enhancer proposes placing both services on the same node, suggesting an \textit{affinity} constraint to minimize carbon emissions.
When such conflicts occur, the Harmonizer resolves them according to the DevOps engineer’s defined priorities: if failure is prioritized, the \textit{affinity} constraint is discarded; if energy is prioritized, the \textit{antiaffinity} constraint is removed. In the absence of an explicit priority, both conflicting constraints are ignored in the subsequent deployment.

\begin{figure}[t]
\footnotesize
\begin{codenum}[firstnumber=1]
    antiaffinity(frontend,large,load_balancer,large).
    #Constraint generated by the Failure Enhancer.
    affinity(frontend,large,load_balancer,large). 
    #Constraint generated by the Energy Enhancer.
\end{codenum}
\caption{Example of conflicting soft constraints generated by the Enhancers.}
\label{fig:harmonizer_example}
\end{figure}

It is important to note that the Harmonizer does not possess a complete view of the MSA. As previously described, its role is limited to verifying that the soft constraints generated by the two enhancers are mutually consistent. If the generated soft constraints conflict with other infrastructure requirements, \eg an avoid constraint applied to the only node capable of hosting a critical component, the Harmonizer is unable to detect or correct such conflicts directly. In these cases, if the \mcs determines that the deployment with all constraints provided by the Harmonizer is unsatisfiable, the problem is re-executed multiple times. A detailed description of the \mcs and how it works is provided in \Cref{subsec:solver}.

\subsection{\mcs}
\label{subsec:solver}

Following the description of the FREEDA framework~\cite{toit} and its model (cf.~\Cref{sec:background}), the application \A and the infrastructure \I are converted into a MiniZinc~\cite{minizinc} data file. MiniZinc is a high-level, solver-independent modeling language for expressing constraint satisfaction and optimization problems. It enables users to define models in a standardized form that can be executed on multiple solvers without modification. Typically, a MiniZinc model is specified in a dedicated model file, while problem-specific data are provided in a separate data file. However, both the model and data can also be combined within a single file if desired.


As described in \Cref{fig:toolchain}, the configuration files \A and \I are updated based on the data collected during previous calls of the Enhancers (\ie from the second call of the \mcs onward, energy values for each component are revised, failed nodes are removed from the configuration, and resource usage and availability are updated accordingly). The new constraints generated by the analyzers after harmonization, together with the updated configuration files, are then re-converted to reflect the changes into a new MiniZinc model. This new model adopts the objective function described in \Cref{eq:newObjFun} to minimize component relocations between nodes, thereby improving deployment stability across successive calls of the solver.

The solver now possesses the full view of the MSA \ie the requirements of each component, the capability of each node and the constraints that have already been harmonized, when present. As mentioned in \Cref{subsec:harmonizer}, the constraints provided by the Harmonizer must be integrated with those defined by the FREEDA model~\cite{toit}. If no additional constraints are present, the solver attempts to compute an optimal deployment according to the criteria defined by the FREEDA model~\cite{toit}, namely maximizing the importance of the flavours assigned to each deployed component. This is typically the case during the initial simulation round, when no prior deployment data are available. If no feasible deployment can be found, the solver returns unsatisfiable and the toolchain terminates, explicitly informing the user that no valid deployment exists for the given application and infrastructure. Conversely, if a feasible deployment is identified, the solution is passed to the next stage of the toolchain as the best deployment plan \D.

However, when additional constraints are provided by the Harmonizer, merging a priori, cannot guarantee that the deployment will be satisfiable, as the interaction between constraints is not easily predictable in the general case. 

Therefore the solver attempts to solve the problem including all constraints to assess whether all of them can be simultaneously satisfied. If a feasible deployment is found, the solver passes it to the next stage of the toolchain. Otherwise, one constraint is removed at a time, and the solver attempts to find a feasible deployment. If no solution is found, pairs of constraints are removed, then triples, and so on, until a deployment is obtained. Whenever a feasible deployment is found, the process stops and the resulting deployment is used as valid for the next simulation round. If no valid deployment can be identified among all combinations, the toolchain terminates and informs the user that no satisfactory deployment was found.

\section{Simulation}
\label{sec:simulation}
This section provides an overview of the entire simulation toolchain, followed by detailed descriptions of each step in the subsequent subsections.

\begin{figure*}[ht]
    \centering
    \includegraphics[width=\textwidth]{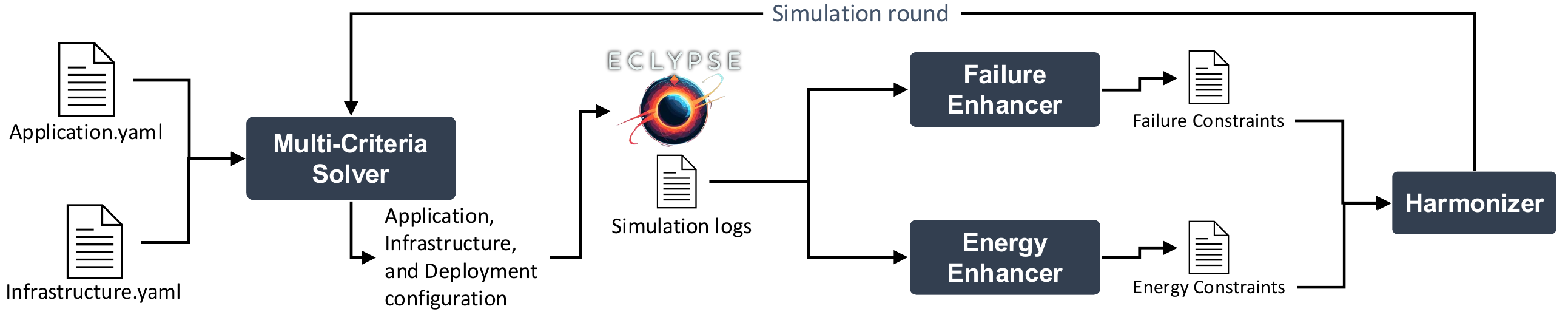}
    \caption{Overview of the toolchain}
    \label{fig:toolchain}
\end{figure*}

As shown in \Cref{fig:toolchain}, the toolchain begins with the \textit{YAML} \cite{YAML} descriptions of the infrastructure and application, which are provided as input to the \mcs. These files are parsed, following the approach described in \cite{toit}, into a MiniZinc data file. The \mcs produces a deployment plan, which is then translated into an \eclypse \cite{eclypse} configuration.

\eclypse simulations are executed in multiple rounds, each combining an infrastructure scenario and an application scenario. Scenarios simulate failures or energy spikes within each simulation round. Application scenarios typically increase the workload or demand on specific components, while infrastructure scenarios reduce the available resources of certain nodes. This approach enables systematic and controlled experimentation across a wide range of operational conditions.

After each round, the simulation logs are processed by both the \fe and the \ee, which generate additional constraints and update the YAML files (both with new energy values and new availability of each infrastructure node) to improve the deployment. These constraints are then passed to the \textit{Harmonizer}, which resolves immediate inconsistencies by prioritizing either failure resilience or energy efficiency, depending on the preference of the user. The harmonized constraints are subsequently fed into the \mcs, which integrates them into the next deployment to adapt to changes in the infrastructure or application imposed by the scenarios. The process is iterative, with each new simulation round executed on the updated deployment.

\begin{figure*}[ht]
    \centering
    \begin{tikzpicture}[
  box/.style={
    rectangle, draw, rounded corners,
    align=center
  },
  arrow/.style={-Latex, thick},
  scale=1, every node/.style={scale=1}
]

\node[box] (lb) {Load balancer};
\node[above=2pt of lb] (lbtext) {\footnotesize{Tiny, Large}};

\node[box, below=of lb] (fe) {Frontend};
\node[below=2pt of fe] (fetext) {\footnotesize{Tiny, Large}};

\node[box, above right=of fe, xshift=1cm] (api) {API};
\node[above=2pt of api] (apitext) {\footnotesize{Tiny, Medium, Large}};

\node[box, below=of api] (redis) {Redis};
\node[below=2pt of redis] (redistext) {\footnotesize{Tiny, Large}};

\node[box, right=2.5cm of api] (idp) {Identity provider};
\node[above=2pt of idp] (idptext) {\footnotesize{Tiny, Large}};

\node[box, below=of idp] (etcd) {etcd};
\node[below=2pt of etcd] (etcdtext) {\footnotesize{Large}};

\node[box, right=of redis] (db) {Database};
\node[below=2pt of db] (dbtext) {\footnotesize{Large}};

\draw[arrow] (lb) -- (fe);
\draw[arrow] (fe) -- (api);
\draw[arrow] (fe) -- (redis);
\draw[arrow] (api) -- (idp);
\draw[arrow] (api) -- (etcd);
\draw[arrow] (api) -- (db);
\draw[arrow] (api) -- (redis);
\draw[arrow] (idp) -- (etcd);

\end{tikzpicture}
    \caption{Overview of the application with flavours available for each component.}
    \label{fig:application}
\end{figure*}
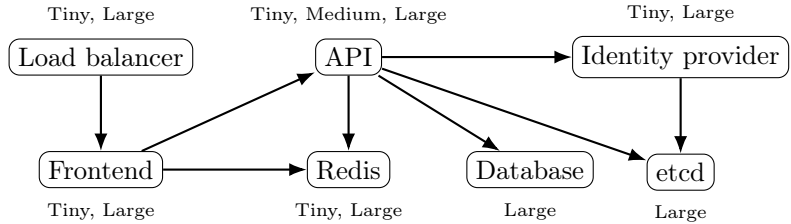

\begin{figure*}[ht]
    \centering
    \begin{tikzpicture}[
  node distance=1.7cm, 
  box/.style={
    rectangle, draw, rounded corners,
    align=center
  },
  arrow/.style={Latex-Latex, thick},
  scale=1, every node/.style={scale=1}
]

\node[box] (pb1) {Public1};
\node[box, below=of pb1] (pb2) {Public2};
\node[box, right=of pb1] (pr1) {Private1};
\node[box, below=of pr1] (pr3) {Private3};
\node[box, right=of pr1] (pr2) {Private2};
\node[box, below=of pr2] (pr5) {Private5};
\node[box, right=of pr2, yshift=-1.1cm] (pr4) {Private4};

\draw[arrow] (pb1) -- (pb2);
\draw[arrow] (pb1) -- (pr1);
\draw[arrow] (pb1) -- (pr3);
\draw[arrow] (pb2) -- (pr3);
\draw[arrow] (pb2) -- (pr1);
\draw[arrow] (pr1) -- (pr2);
\draw[arrow] (pr1) -- (pr5);
\draw[arrow] (pr1) -- (pr3);
\draw[arrow] (pr1) -- (pr4);
\draw[arrow] (pr3) -- (pr2);
\draw[arrow] (pr3) -- (pr5);
\draw[arrow] (pr3) -- (pr4);
\draw[arrow] (pr2) -- (pr5);
\draw[arrow] (pr4) -- (pr5);
\draw[arrow] (pr4) -- (pr2);

\end{tikzpicture}
    \caption{Infrastructure node with connections}
    \label{fig:infrastructure}
\end{figure*}
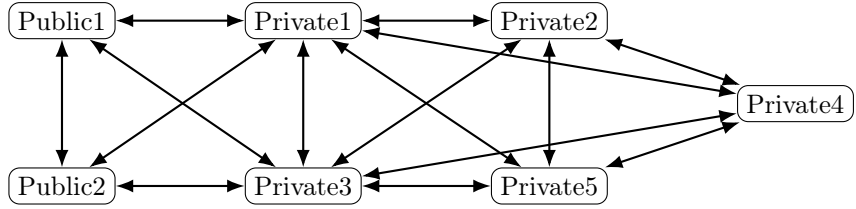

\subsection{Simulation Setup}
\smallskip \noindent
\paragraph{Application and Infrastructure Description}
\label{subsec:appinfrdescription}
To run the simulation, we consider a reference architecture, illustrated in
\Cref{fig:application}. In the figure, nodes represent the application
components, arrows indicate the dependencies of each component, and the
available flavours of each component are shown above each node. All components
are conceptualized as services that can be deployed on different machines. The
application comprises several components and is representative of a large class
of MSAs~\cite{FLM19}. Specifically, the \emph{Load balancer} distributes
incoming client requests evenly across multiple instances of the frontend
service. The \emph{Frontend} interfaces with users, forwarding requests to
backend services via the API service. In turn, the \emph{API} service acts as
the central business logic hub, processing requests from the frontend, handling
core application functionalities, and orchestrating calls to other backend
components. Among these, \emph{Redis} works as a caching layer, storing
transient data for fast access, useful to accelerate response times and offload
frequent database queries, while the \emph{Database} manages persistent data
storage required by the application. The last two components are the
\emph{Identity provider}, which handles user authentication and authorization,
while \emph{etcd} manages distributed configuration and service discovery.
Flavour-wise, the Load Balancer, Frontend, Redis, and Identity Provider can be
deployed in either Tiny or Large flavours; etcd and the Database are available
only in the Large flavour; and the API component is offered in three flavours:
Tiny, Medium, and Large. The application specification is defined through a YAML
\cite{YAML} file, following the format described in \cite{toit}. The full
description of the application is available in a companion repository
\cite{repo}, defined using the FREEDA YAML specification~\cite{freedayamlspec}.

Similarly to the reference architecture, we define a representative
infrastructure for the simulation. Specifically, we consider an Edge-Cloud
infrastructure with a highly interconnected mesh topology with multiple
redundant paths between nodes, providing resilience and load distribution
capabilities~\cite{ACRC19}. The infrastructure is illustrated in
\Cref{fig:infrastructure}, where nodes represent the physical nodes where
components can be deployed. Nodes can be Public or Private, since certain
services can only be deployed in one of those two categories. Arrows indicate
the physical connections of each node, two mutually dependent components will
need to be deployed on two physically connected nodes. In the infrastructure, we
find nodes \emph{Public1} and \emph{Public2}, which are connected Cloud nodes,
enabling direct within-cloud communication. The \emph{Private} nodes (5 in
total) form a complete subnetwork of connected on-premises edge devices. Note
that the only way for the Public nodes to directly communicate with services
deployed in the on-premises Private subgroup is through nodes \emph{Private1}
and \emph{Private2}.

Furthermore, each node has a \emph{subnet} attribute so that only the services with a matching set of attributes can only be deployed
on the corresponding nodes, \eg each node has in the \emph{subnet} attribute either the value \emph{private} or \emph{public} (depending on the type of node) and each component has this attribute too so it can be deployed only on certain nodes; full description of the app in available at \cite{repo}.
For space reasons, we delegate the full description of these attributes to the
supplementary material found in the companion repository \cite{repo}.

\paragraph{\mcs} 
As illustrated in \Cref{fig:toolchain}, the \mcs receives as input the YAML specifications of the application and infrastructure---either with their initial values or as updated by the \fe and \ee after each simulation round---together with the constraints generated by the harmonizer. At first, the YAML \cite{YAML} specifications files go through a parsing phase to obtain a MiniZinc \cite{minizinc} (version 2.8.5) representation and executed with the base constraints provided in \cite{toit}. Full details of parsing from high-level YAML specification to the low-level Minizinc one can be found in \cite{toit}.

Following what is specified in \Cref{subsec:solver}, the solver (selected from those available within the MiniZinc distribution and defaulting to Gecode 6.3.0 also used in our experiments) aims to produce an optimal deployment whenever one exists within the time limit of 5 minutes, according to the criteria that have been chosen based on the simulation round (\ie maximizing the importance of each flavour in the first round of the simulation or relocating the least amount of components in subsequent rounds).
Once the computation is complete, the resulting deployment is written to a text file that is subsequently parsed into a \eclypse simulation object.


\paragraph{\fe}
This component takes as input the simulation logs generated from \eclypse and, as described in \Cref{subsec:failure-enhancer}, generates constraints of type \textit{Affinity}, \textit{Anti-affinity} or \textit{Avoid}.
\Cref{fig:toolchain} illustrates the workflow of the \fe during the simulation. The process begins with the \textit{Simulation.log} file generated by the \eclypse simulator, a sample of which is shown in \Cref{fig:simulationLog}. This log is processed by a component called \eclypse Parser, which extracts the relevant information and generates the Prolog facts required by the \fe. These facts describe both deployment and failure events, covering components as well as nodes, as depicted in \Cref{fig:deploymentPL}. Based on this input, the \fe produces deployment soft constraints designed to improve the resilience of the current MSA deployment, which are subsequently passed to the Harmonizer for processing.

\begin{figure}[t]
\footnotesize
\begin{codenum}[firstnumber=1]
    17:20:33|ECLYPSE|Simulation - Event Start-0 fired.
    17:20:33|ECLYPSE|Simulation - Event Tick-0 fired.
    17:20:33|ECLYPSE|Simulation - Event Enact-0 fired.
    17:20:33|ECLYPSE|PlacementManager - 
        Placement of case_study on case_study
    17:20:33|ECLYPSE|PlacementManager - 
        {load_balancer_large -> public1 | 
        api_large -> private1 |
        frontend_large -> public1 | 
        redis_large -> private3 | 
        identity_provider_large -> private3 | 
        database_large -> private5 | 
        etcd_large -> private1}
    17:20:33|ECLYPSE|Simulation - Event Tick-1 fired.
    17:20:33|ECLYPSE|Simulation - Event Enact-1 fired.
    17:20:33|ECLYPSE|PlacementManager - 
        Placement of case_study on case_study
    17:20:33|ECLYPSE|PlacementManager - 
        {load_balancer_large -> public1 | 
        api_large -> private1 | 
        frontend_large -> public1 | 
        redis_large -> private3 | 
        identity_provider_large -> private3 | 
        database_large -> private5 | 
        etcd_large -> private1}
    ...
\end{codenum}
\caption{Extract of a simulation.log file.}
\label{fig:simulationLog}
\end{figure}

\begin{figure}[t]
\footnotesize
\begin{codenum}[firstnumber=1]
    deployedTo(api, large, private1).
    deployedTo(database, large, private5).
    deployedTo(etcd, large, private1).
    ...

    unreachable(frontend, 31).
    unreachable(frontend, 32).
    ...

    overload(public1, cpu, 31, 98).
    ...
\end{codenum}
\caption{Extract of the deployment.pl file generated by the \eclypse Parser.}
\label{fig:deploymentPL}
\end{figure}

\paragraph{\ee}
The \ee takes as input the Infrastructure description \I, where all the information related to the nodes are stored, the Application description \A, where we can find the information about the services and the application Deployment \D, detailing where each service was deployed on which node.
Finally it uses the simulation logs from \eclypse, providing the data about the services and services connections consumptions.

Starting from the reports generated from \eclypse, the \ee reads them and estimates their emissions, this estimate takes into account only the periods of time where each service or node was active and uses the average consumption during the monitored period. These emissions are then passed along inside the \ee to generate the constraints pertaining to the current deployment that was measured. 
The constraints generated can be of the type \textit{Affinity} or \textit{Avoid}.
An \textit{Affinity} constraint indicates that two services exchange a lot of data and thus should be paired together on the same node.
An \textit{Avoid} constraint indicates that a service would consume too much if put on the node targeted by this constraint, and thus it would be best to deploy it somewhere else.
Once the constraints are generated, we save them inside a \textit{Knowledge Base} so that in the future we can check the recurrence of the generated constraints, potentially indicating that they are relevant and thus should be proposed again.

Once the \ee completes its execution, it will provide in output the \textit{Energy Constraints} automatically ranked by estimated emissions produced and assigned a weight indicating the internal ranking. An example of its output can be seen in Figure \ref{fig:energyconstraintsexample}, where the constraint type is followed by the elements it is referring to, and finally the weight associated to them.

\begin{figure}[t]
\footnotesize
\begin{codenum}[firstnumber=1]
    avoid(d(identity_provider,large),private3,0.883).
    avoid(d(database,large),private1,1.0).
\end{codenum}
\caption{Example of Energy Related constraints.}
\label{fig:energyconstraintsexample}
\end{figure}

\paragraph{Harmonizer}
As shown in \Cref{fig:toolchain}, the Harmonizer takes as input the soft constraints generated by both the \fe and the \ee. Its role is to identify potential conflicts among these constraints and resolve them according to the user-defined priority, which may emphasize resilience, sustainability, or a balance between them. Once the constraints have been processed, they are sent to the \mcs to be considered for the next deployment.

The infrastructure is depicted in \Cref{fig:infrastructure}, consisting of seven machines represented as nodes. These are divided into 2 public nodes (meant to be publicly exposed) and 5 private nodes (intended to simulate nodes within a private subnet). As detailed in the repository \cite{repo}, each node includes a resource list attribute called subnet, which specifies whether the node is exposed to the internet. Specifically, nodes Public1 and Public2 are assigned the value public, while the remaining private nodes are assigned the value private. Blue (bi-directional) arrows between nodes represent connections between machines.
A complete view of the application, described in YAML format, can be seen at \cite{repo}, following the specification described in \cite{toit,freedayamlspec}.

\subsection{Scenarios}
In this section, we describe the scenarios evaluated against both the application and the underlying infrastructure, outlining the expected results and the observed outcomes.

A scenario is defined as a set of rules that modify the behaviour of elements during a simulation round. These rules may affect infrastructure components, such as CPU or RAM availability on a node, or application-level aspects, such as the energy consumption of a service. Rules can also be applied selectively to specific time intervals, referred to as simulation ticks, providing flexibility in the types of experiments that can be conducted.
During the experimentation phase, two recurring patterns were employed:
\begin{itemize}
    \item \textit{Constant modification}, where a resource is adjusted by a fixed value over a defined range of ticks.
    \item \textit{Sinusoidal modification}, where a resource varies following a sinusoidal function, ensuring that the starting and ending values align.
\end{itemize}

The first pattern was used to stress test thresholds at which a deployment might lack sufficient CPU or RAM to host a service, thereby inducing failures and downtime within the system. The second pattern was designed to simulate abnormal but transient behaviours, \eg where no issues are evident at the start or end values, showcasing how the FREEDA approach is able to detect such instances and manage them accordingly.

The \eclypse simulator incorporates a mechanism referred to as Update Policy, which enables controlled environmental dynamism by programmatically modifying both infrastructure resource capacities and application-level requirements across simulation rounds. Within the experimentation phase, this mechanism was employed to introduce predefined scenarios at specific points in the simulation timeline. \Cref{fig:update_policy} presents an example of the configuration schema used for this purpose. In this structure, the position within the array specifies the simulation round at which a particular policy is applied, as well as the system component it targets, either the infrastructure or the application. This flexible approach enables not only the execution of multiple scenarios but also the exploration of various combinations of scenarios within a single simulation.
Within the project repository, you can find the results of several simulations combining multiple scenarios \cite{case_study_experiments}.

\begin{figure}[t]
\footnotesize
\begin{codenum}[firstnumber=1]
    application_policies = [
        Example 1: [scenario] * X  # we run this 
            scenario X times on the application side
        Example 2: [scenario1] * i + [scenario2] * j 
            # we run scenario1 i times and then we run 
            scenario2 j times on the application side.
    ]
    infrastructure_policies = [
        Example 1: [scenario] * Y # we run this 
            scenario Y times on the infrastructure side
        Example 2: [scenario3] * n + [scenario4] * m 
            # we run scenario3 n times and then we run 
            scenario4 m times on the infrastructure side.
    ]
\end{codenum}
\caption{Code snippet showing the configuration schema used to run the scenarios.}
\label{fig:update_policy}
\end{figure}

An energy malfunction from a service does not directly result in downtime from said service, but rather in worse green performances, impacting the total emissions. For this reason, the energy scenarios can be flexibly applied to any service of the MSA, so that the different resulting deployments can be easily explored.

The \eclypse simulator also provides integrated deployment strategies for placing applications onto nodes. Under the \emph{first-fit} strategy, services are placed on the first available node that satisfies their resource requirements. In contrast, the \emph{best-fit} strategy selects the node that maximizes resource utilization without exceeding capacity. The goal of the defined scenarios and simulations is to compare outcomes obtained using the integrated \eclypse deployment strategies against those achieved with the FREEDA approach.

The primary objective of this scenario is to degrade the CPU and RAM using a \textit{constant modification} of the resources available for the public nodes, thereby compromising the operation and deployment of the Load-Balancer and Frontend services.
When this degradation of resources occurs, the affected services are expected either to be relocated to a different node or to be redeployed with a different flavour than the one originally assigned. This setup allows us to evaluate how both the \eclypse \emph{best-fit} placement strategy and the FREEDA approach respond to resource constraints on essential nodes and whether they can adapt by reallocating or reconfiguring services to maintain the MSA availability. The results obtained for this scenario are discussed and analyzed in \Cref{sec:results}.

\subsection{Results}
\label{sec:results}

As showcased in Figure \ref{fig:update_policy} the scenarios applied during the simulation phase aimed to tackle a situation where certain nodes degraded their CPU and RAM resources through a \textit{constant modification}, while some services degraded their energy performances through a \textit{sinusoidal modification}.

During the simulation process, we evaluated five different configurations of the toolchain to observe how these components behave under identical scenarios. These toolchain configurations are summarized below:

\begin{enumerate}
\item \textbf{\eclypse \emph{best-fit}}: The application is deployed using only the placement strategy \textit{best-fit} provided by the simulator.

\item \textbf{Only \mcs}: The application is deployed using the deployment configuration generated by the \mcs.

\item \textbf{\mcs and \ee}: The application is deployed using the \mcs's deployment configuration, enriched with constraints generated by the \ee based on simulation logs.

\item \textbf{\mcs and \fe}: The application is deployed using the \mcs's deployment configuration, combined with constraints generated by the \fe from the simulation logs.

\item \textbf{Full FREEDA}: The application is deployed using the complete FREEDA toolchain, \ie the \mcs's deployment configuration together with the constraints generated by both the \ee and \fe.
\end{enumerate}

\begin{figure*}[p]
    \centering
    \includegraphics[width=\textwidth, height=\textheight, keepaspectratio]{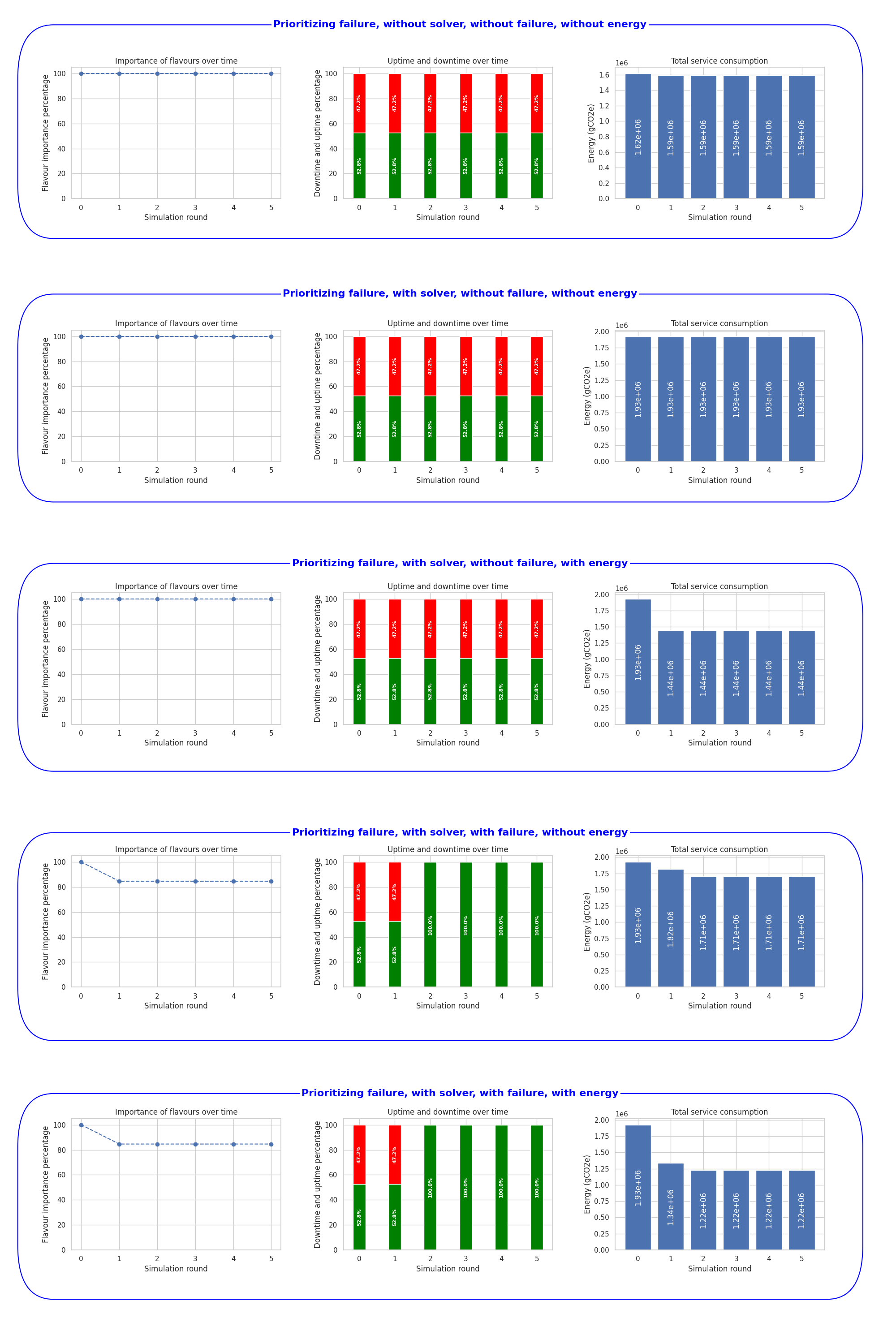}
    \caption{Results obtained for our running example for each toolchain configuration, with priority failure.}
    \label{fig:general_plot}
\end{figure*}

The first toolchain configuration (\eclypse \emph{best-fit}) relies on the internal \emph{best-fit} strategy of the \eclypse~\cite{eclypse} simulator, deploying all components in the Large flavour, \ie the most resource-intensive configuration. This approach is intended to emulate the decisions of an expert DevOps engineer who is able to deploy the full version of the application with full knowledge of both the application requirements and the infrastructure capabilities. In this setting, all seven components were successfully deployed across all six simulation rounds. During the second deployment (corresponding to round 1), \eclypse reallocated certain components to reduce energy consumption, after which it maintained this configuration, resulting in stable energy usage. With respect to failures, since the scenario is repeated, the downtime plot exhibits a consistent pattern of failures across rounds.

The second row of~\Cref{fig:general_plot} illustrates the second toolchain configuration (Only \mcs). As in the previous case, the \mcs attempts to deploy all components in their most powerful flavour. However, since the \mcs does not consider energy optimization, the energy consumption is higher than in the first methodology. However, because the deployed configuration remains unchanged across all simulation rounds, energy usage remains stable throughout. The failure patterns are identical to those observed previously, as the \mcs alone (without the \fe and \ee) lacks awareness of the events occurring during the simulation rounds.

The \ee is the module in charge of energy emissions from the deployment as a whole. The module is active in the third and fifth rows of Figure \ref{fig:general_plot}. During the simulation it was put in place a scenario negatively impacting the database, causing a spike in energy consumed from that service. Below in Figure \ref{fig:EE_output_round0} we can see which constraints were generated from the \ee after the first simulation round. The database is correctly identified as the biggest energy consumer, and this is also reflected on the weights assigned, with the biggest priority being the database. Since the database requires an encrypted storage as a security measure for it to be deployed on a node, and being there only two nodes which such feature in the infrastructure used as an example, the \ee will correctly propose soft constraints only for one of those two nodes, since we want to deploy the database service somewhere.

\begin{figure}[t]
\footnotesize
\begin{codenum}[firstnumber=1]
    avoid(d(identity_provider,large),private1,0.493).
    avoid(d(identity_provider,large),private3,0.883).
    avoid(d(identity_provider,large),private5,0.413).
    avoid(d(database,large),private1,1.0).
\end{codenum}
\caption{Soft constraints generated by the \ee during the first simulation round.}
\label{fig:EE_output_round0}
\end{figure}

For all the subsequent rounds the constraints generated from the \ee remain the same, causing the energy consumptions of the services to stabilise, as can be seen in Figure \ref{fig:general_plot}.
The same behaviours are mirrored in the 5th line, where both the \ee and the \fe are considered.

The \mcs and \fe configuration is illustrated in the third row of \Cref{fig:general_plot}. In this setup, the constraints generated by the \fe enable the \mcs to maintain 100\% uptime, even under resource degradation on the Public1 and Public2 nodes.
\Cref{fig:deploymentPL_round0} shows the input to the \fe for the first simulation round, as generated by the \eclypse Parser from the simulation logs. In this round, both the Frontend and Load-Balancer services were deployed on Public1 using the Large flavour. Since the \emph{best-fit} scenario deliberately degrades the resources of Public1, both services become unreachable between Ticks 31 and 98, as expected. Based on this information, the \fe produces the soft constraints shown in \Cref{fig:FE_output_round0}, which are then passed to the Harmonizer and subsequently processed by the \mcs.

\begin{figure}[t]
\footnotesize
\begin{codenum}[firstnumber=1]
    deployedTo(api, large, private1).
    deployedTo(database, large, private5).
    deployedTo(etcd, large, private1).
    deployedTo(frontend, large, public1).
    deployedTo(identity_provider, large, private3).
    deployedTo(load_balancer, large, public1).
    deployedTo(redis, large, private3).
    
    unreachable(frontend, 31)...
    ...unreachable(frontend, 98).
    unreachable(load_balancer, 31)...
    ...unreachable(load_balancer, 98).
    
    overload(public1, cpu, 31, 98).
    overload(public1, ram, 31, 98).
\end{codenum}
\caption{Deployment.pl file generated by the \fe during the first simulation round.}
\label{fig:deploymentPL_round0}
\end{figure}

\begin{figure}[t]
\footnotesize
\begin{codenum}[firstnumber=1]
    avoid(d(frontend,large),public1).
    avoid(d(load_balancer,large),public1).
\end{codenum}
\caption{Soft constraints generated by the \fe during the first simulation round.}
\label{fig:FE_output_round0}
\end{figure}

\subsection{Discussion}

Observing Figure \ref{fig:general_plot}, we can see how the full FREEDA approach is able to leverage all the improvements suggested by each single module and incorporate them in a single solution that is both failure resilient and energy observant at the same time.
At the start we can notice how the native \eclypse deployment strategy might achieve slightly better energy results than the first FREEDA toolchain configuration, but we must keep in mind that the \eclypse best fit does not account for the various deployment configurations needed, as described in Paragraph \ref{subsec:appinfrdescription}, it simply fits the most consuming service in the greenest node, without properly checking if the service requires a private or a public node or the service dependencies, or other requirements, such as the encrypted storage for the database service.
This leads to the initial best fit deployment resulting greener than the FREEDA configurations without the \ee component.
From the Failure standpoint, the \eclypse best fit does not have ways to resolve problems that might arise, leading to continuous service disruptions.
In the last row, the full FREEDA approach is utilised. At first, the energy consumption might result in higher values, due to the reasons explained previously, but they are swiftly brought down, below the initial deployment and even the \eclypse deployment strategy. This can be attributed to energy constraints being observed from the \mcs and the \mcs deciding to change the flavour of a component, thus diminishing the flavour importance too. From the second simulation round onwards, the situation stabilizes, with a successful maximization of flavour importance, while removing entirely failures and minimizing the energy consumption.

\section{Emulation}
\label{sec:emulation}
This section provides an overview of the emulation toolchain, followed by detailed descriptions of each step in the subsequent subsections.
The primary objective of the emulation phase is to gather data on the resilience and carbon efficiency provided by FREEDA. Specifically, we measure service availability, application quality, and the carbon emissions produced by the cluster.

All emulation experiments were executed on a virtualized Kubernetes cluster deployed using Minikube, enabling reproducible test scenarios under controlled conditions. This setup allows for the intentional introduction of performance degradation, adjustments in energy consumption parameters to emulate carbon intensity fluctuations, and the imposition of computing resource constraints on infrastructure nodes.

\subsection{Emulation Setup}
\paragraph{Reference Application}
To evaluate the FREEDA toolchain, described in \Cref{sec:toolchain}, no existing benchmark was available that supported the management of multiple service flavours. Therefore, a new application called BrewMonitor was designed and implemented. BrewMonitor is an open-source, flavoured MSA, publicly available on GitHub at \cite{brew-monitor}. It follows a cloud-native architecture composed of several independent microservices that can be configured in two flavours (Tiny and Large) and simulates the collection, aggregation, and analysis of data within a realistic environment.

\begin{figure}[t]
    \centering
    \includegraphics[trim=2cm 0cm 12cm 0cm,width=\columnwidth]{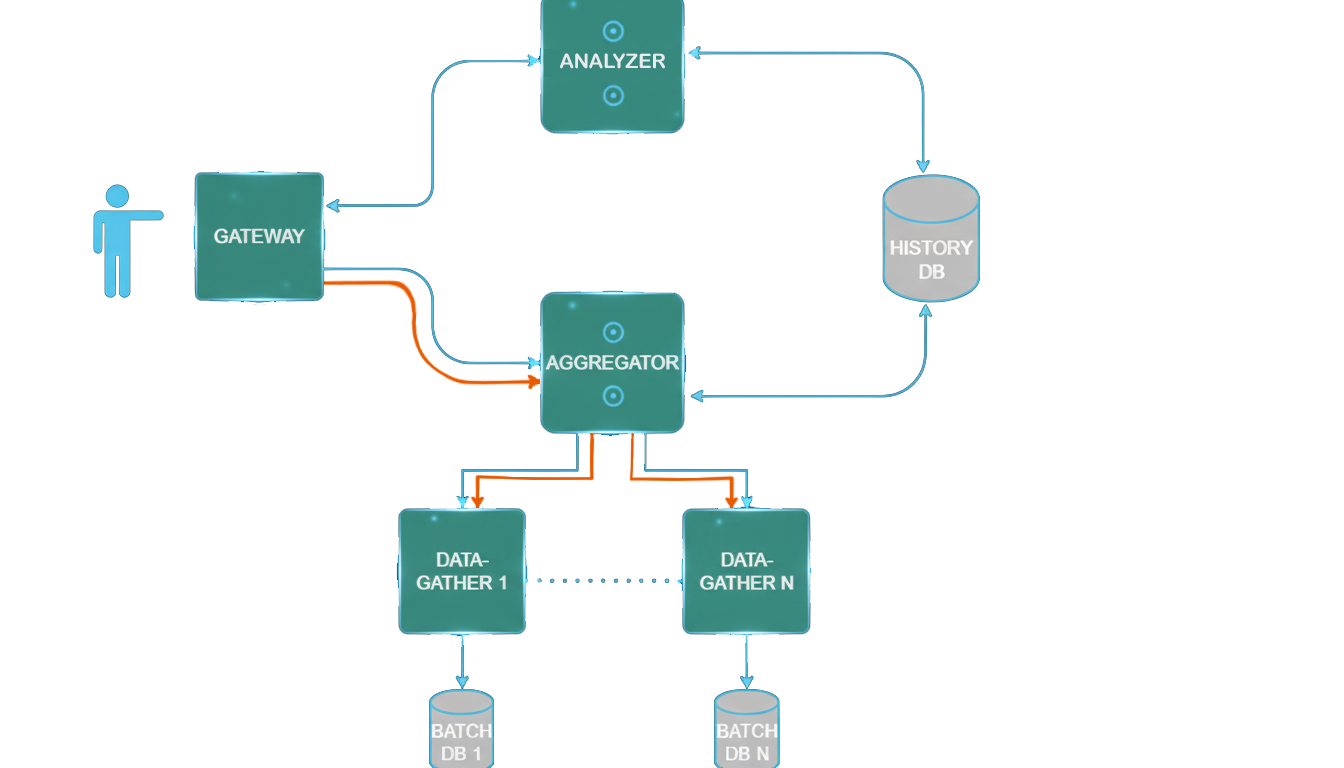}
    \caption{Overview of the BrewMonitor MSA architecture.}
    \label{fig:brewmonitor}
\end{figure}

\Cref{fig:brewmonitor} illustrates the architecture of our reference application. BrewMonitor is a monitoring application for brewing plants, implemented as an MSA designed for modularity, scalability, and ease of extension. Its primary purpose is to collect real-time data from brewery sensors, aggregate and analyze it, and expose the results through REST interfaces for production operators and supervisory systems. The architecture was designed to support two distinct \emph{flavours}: \textit{Tiny} (red arrows in \Cref{fig:brewmonitor}) and \textit{Large} (light blue arrows), which differ in functionality, data persistence, and resilience features. The MSA is composed of four main microservices, each responsible for a specific stage of the monitoring workflow:

\begin{itemize}
    \item \textbf{Gateway:} Serves as the single entry point to the system, exposing HTTP REST endpoints that unify access to the underlying services. Implemented in Python using Flask, it provides lightweight integration, centralized logging, authentication, and rate limiting.
    \item \textbf{Data Gather:} Handles the sampling of process data (temperature, humidity, and pH) from simulated sensors. In the \textit{tiny} flavour, values are generated in memory upon request, providing a lightweight, stateless service. In the \textit{Large} flavour, readings are persisted in MongoDB with a 24-hour TTL, enabling access to historical data for analysis and auditing.
    \item \textbf{Aggregator}: Periodically collects data produced by one or more \textit{Data Gather} instances. In the \textit{Tiny} flavour, it queries the \textit{/data-gather/avg} endpoint and forwards the results directly to clients. In the \textit{Large} flavour, a background thread polls each \textit{Data Gather} instance every minute, stores the data in MongoDB, and exposes the latest aggregated records via the \textit{/aggregator/current} endpoint, ensuring consistency across collection nodes.
    \item \textbf{Analyzer:} Available only in \textit{Large} flavour, this component performs statistical analyses on historical MongoDB data. It computes metrics such as maximum, minimum, standard deviation, and outliers for temperature, humidity, and pH, and exposes a single \textit{/analyzer/stats} endpoint that returns detailed JSON results for each monitored instance.
\end{itemize}

\paragraph{Cluster and node configuration}
The experimental environment consists of a Minikube cluster composed of seven virtual nodes, each with distinct technical specifications reflecting the heterogeneity typical of modern hybrid cloud environments.

\begin{itemize}
    \item \textit{gatewaynodeefficient}: designated as the cluster master, features a balanced configuration with 2 CPU cores and 4 GB of RAM, and a carbon intensity of 200 gCO2/kWh. This setup is ideal for hosting gateway services that prioritize stability over raw performance.
    \item \textit{gatewaynodestrong}: provides enhanced computational capacity with 4 CPU cores and 8 GB of RAM but exhibits a slightly higher carbon intensity of 250 gCO2/kWh, representing the classic trade-off between performance and sustainability.
    \item \textit{databasenode}: serves as the primary node for data management, equipped with 8 CPU cores and 8 GB of RAM and optimized for a low carbon intensity of 100 gCO2/kWh. This configuration underscores the importance of achieving high performance in data persistence while maintaining a reduced environmental footprint.
    \item \textit{appnodestrong}: serves as the main node for executing core application logic, equipped with 8 CPU cores, 8 GB of RAM, and optimized for a low carbon intensity of 100 gCO2/kWh.
    \item \textit{appnodeefficient}: offers a balanced compromise, with 6 CPU cores and 6 GB of RAM, but stands out as the most carbon-efficient node in the cluster, with a carbon intensity of only 90 gCO2/kWh. This node is particularly useful for testing the FREEDA toolchain's ability to prioritize carbon efficiency under favourable conditions.
    \item \textit{veryexpensive}: includes 8 CPU cores and 8 GB of RAM but is intentionally configured with a high carbon intensity of 1000 gCO2/kWh, simulating emergency scaling scenarios on environmentally costly infrastructure.
\end{itemize}


\paragraph{Metrics and Monitoring}
To monitor the reference MSA, a custom observation script was developed. This script adopts a multi-layered architecture for collecting and processing application metrics. By integrating native Kubernetes APIs, the Prometheus monitoring system, and Kepler (for energy consumption measurement), it provides a comprehensive view of the cluster's performance at both the functional and energy levels.

During the emulation, three key metrics were monitored: \textit{App Quality}, \textit{Downtime}, and \textit{C02 Emissions}.
App Quality represents an aggregate measure of the importance of the flavours currently active within the MSA. It is expressed as the percentage ratio between the current importance score and the maximum theoretically achievable. This metric provides a quantitative indication of FREEDA's ability to maintain services running on the highest-performing flavours.

The Downtime Percentage quantifies the proportion of time during which the MSA is unable to deliver all required services at the minimum acceptable quality level. It is computed by continuously monitoring the readiness status of all application pods, excluding auxiliary pods (\eg ballast pods used to simulate system stress). A period is considered uptime only when all expected pods are simultaneously in the Ready state, providing a realistic measure of overall service availability.

The total CO2 emissions constitute the primary metric for assessing the environmental impact of the MSA. They are calculated by integrating over time the product of instantaneous power consumption and the carbon intensity associated with each active node. Electrical power data are obtained via Prometheus queries, while carbon intensity values are retrieved from dynamic labels applied to the corresponding Kubernetes nodes.

\subsection{Scenarios}
The experimental suite defined for the emulation comprises seven scenarios across four thematic series, each targeting specific adaptive features of the FREEDA toolchain and its multi-objective optimization mechanisms.

\begin{itemize}
    \item \textbf{Series 1.x - Resource Exhaustion:} Tests FREEDA's handling of resource scarcity. Scenarios 1.1 and 1.2 artificially increase CPU demands for critical services, triggering rescheduling and activating failure-management logic.
    \item \textbf{Series 2.x - Nodal Stress and Carbon Variation:} Evaluates FREEDA's response to localized contention and environmental changes. Scenario 2.1 simulates heavy resource usage via ballast pods, while Scenario 2.2 dynamically raises node carbon intensity to test carbon-aware workload rebalancing.
    \item \textbf{Series 3.x - Failures and Policy Dilemmas:} Explores compound faults and trade-offs. Scenario 3.1 combines node failure with resource overload, and Scenario 3.2 introduces multiple stresses and conflicting optimization goals to assess FREEDA's decision-making under complex conditions.
    \item \textbf{Series 4.x - Carbon-Aware Adaptation:} Examines continuous adaptation to changing environments. Sub-scenarios 4.1 to 4.5 progressively vary carbon intensity, workload, and node capacity to test FREEDA's responsiveness to gradual and sudden shifts in both environmental and infrastructural parameters.
\end{itemize}

This experimental suite provides a comprehensive assessment of FREEDA's resilience, efficiency, and environmental adaptability under diverse and evolving operational conditions.
The comparison methodology involves executing each scenario sequentially under two distinct configurations. The baseline phase relies solely on native Kubernetes orchestration, with no intervention from FREEDA. The adaptive phase then activates FREEDA's optimization mechanisms to mitigate the effects of the introduced degradations. This design enables a direct and quantitative comparison between the two configurations, isolating the influence of FREEDA's algorithmic decisions and allowing observed improvements to be confidently attributed to the framework's adaptive strategies.

\begin{figure}[t]
    \centering
    \includegraphics[width=\columnwidth]{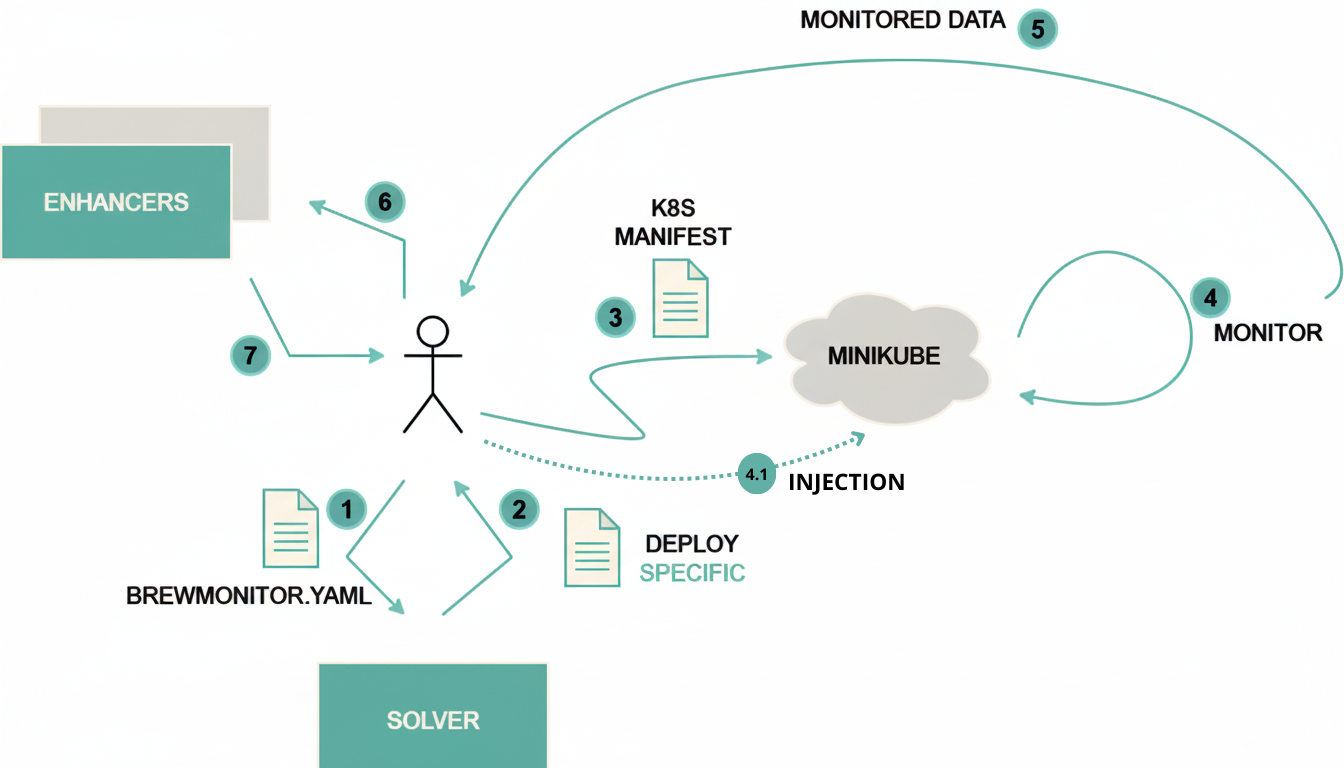}
    \caption{Workflow followed during the execution of each emulated scenario.}
    \label{fig:emulation-workflow}
\end{figure}

Whereas the baseline scenarios rely on native Kubernetes orchestration, in contrast, the scenarios optimized with FREEDA implement a manual feedback loop to simulate an intelligent control system, as illustrated in \Cref{fig:emulation-workflow}.
Each scenario is structured to first expose the limitations of traditional orchestration, documenting the failure patterns, energy inefficiencies, and performance bottlenecks that arise in the absence of intelligent optimization. The subsequent analysis focuses on the corrective actions implemented by FREEDA, \eg service migration between nodes, flavour adaptation, or overall resource reallocation, demonstrating how these mechanisms enhance resilience, efficiency, and sustainability across the cluster.

\subsection{Results}
In this section, we present the results for scenarios 2.2 and 3.1. Results for all other scenarios, along with their execution instructions, are publicly available on GitHub at \cite{emulation-results}.

\begin{figure}[t]
    \centering
    \includegraphics[width=\columnwidth]{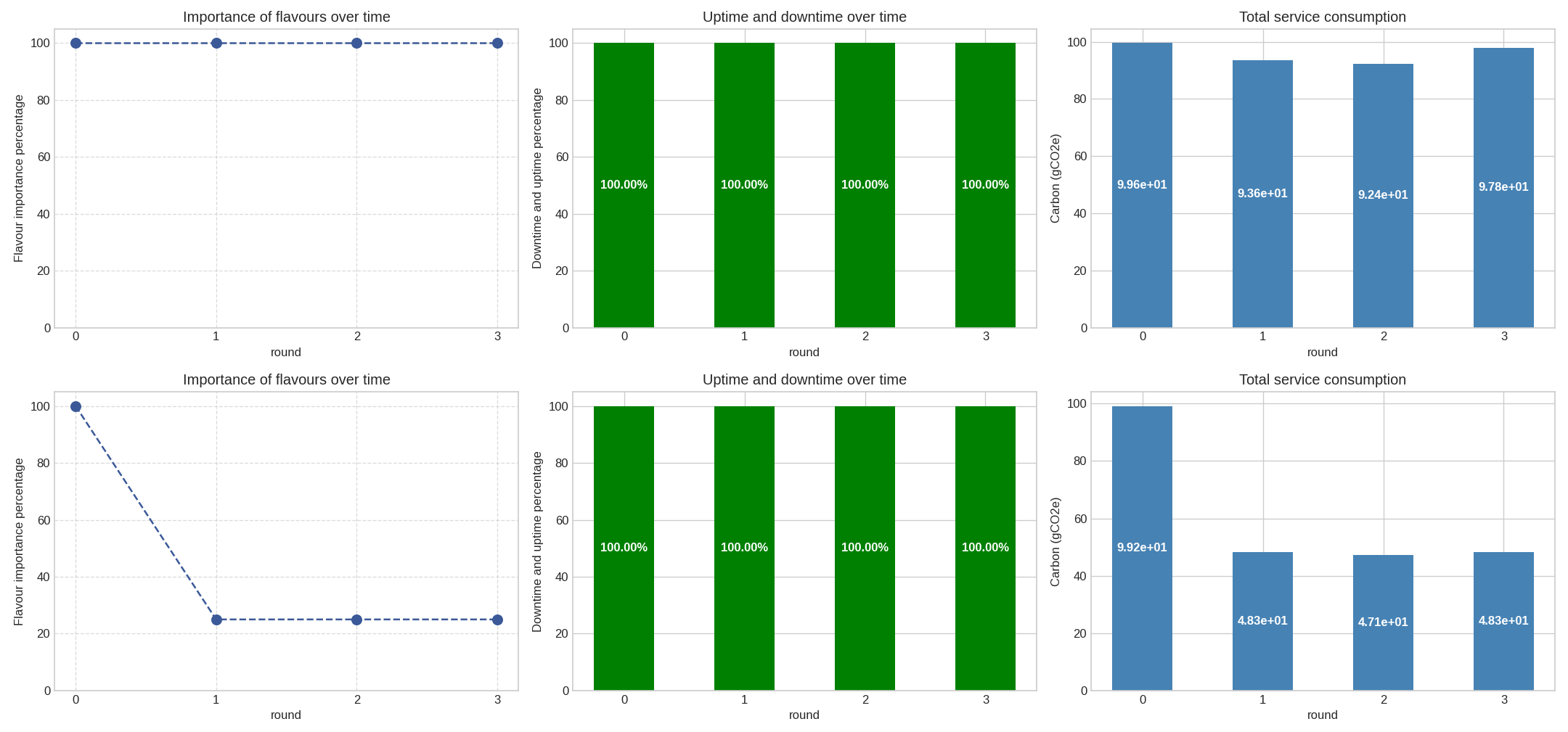}
    \caption{Results of the emulation of scenario 2.2, the first row corresponds to the baseline results, and the second row to those obtained using FREEDA.}
    \label{fig:scenario2-2}
\end{figure}

During scenario 2.2, we implement a dynamic variation of the carbon intensity of the \textit{appnodestrong} node, increasing it from 100 to 700 gCO2/kWh. This scenario evaluates FREEDA's carbon-aware capabilities, verifying if the framework is able to detect changes in environmental conditions and consequently rebalance workload placement to minimize the overall carbon impact.
A detailed comparison between the baseline execution and the FREEDA-optimized deployment is illustrated in \Cref{fig:scenario2-2}. As shown, FREEDA's intervention fundamentally transforms the environmental profile of the MSA, demonstrating the potential of carbon-aware optimization strategies for modern MSAs.

The lower left panel of \cref{fig:scenario2-2} shows that \textit{App Quality} experiences an immediate, controlled reduction to approximately 25\% from the first optimization round, remaining stable thereafter.
This controlled reduction indicates that FREEDA implemented an aggressive but calibrated re-deployment strategy. It involved a large-scale migration of services from the high-carbon-intensity node to significantly more sustainable alternatives and the selective degradation of some flavours to optimize the overall MSA carbon efficiency. This strategy respects the trade-off between the carbon emissions budget and cost. 

Throughout the experiment, \textit{Uptime} (lower central panel) remains at 100\%, demonstrating that FREEDA's carbon-aware optimizations do not compromise operational availability. Overall, the results confirm that FREEDA effectively balances sustainability and service continuity under dynamically changing environmental conditions.

\begin{figure}[t]
    \centering
    \includegraphics[width=\columnwidth]{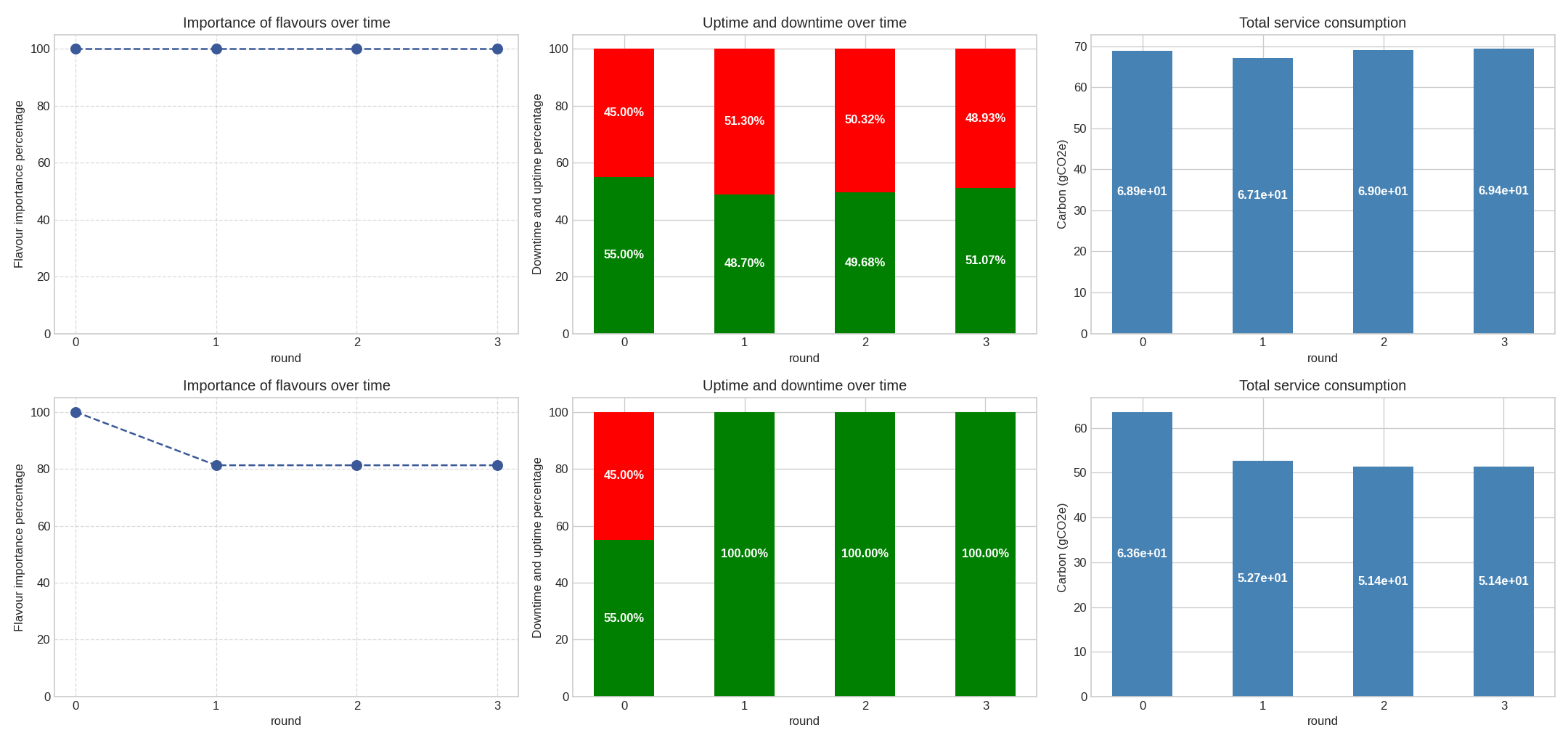}
    \caption{Results of the emulation of scenario 3.1, the first row corresponds to the baseline results, and the second row to those obtained using FREEDA.}
    \label{fig:scenario3-1}
\end{figure}

Scenario 3.1 introduces a multiple-stress condition: first, it simulates the complete failure of the \textit{gatewaynodestrong} node via drain, forcing the migration of all hosted pods; then, it subjects the \textit{aggregator} service in its \textit{Tiny} flavour to a resource overload. The goal is to test FREEDA's capacity to balance fault tolerance, performance, and environmental sustainability under critical stress conditions.
A detailed comparison of the metrics between the baseline approach and the FREEDA-optimized solution for scenario 3.1 can be observed in \Cref{fig:scenario3-1}.

The baseline execution of this scenario (\Cref{fig:scenario3-1}, top row) highlights the severe impact of a node failure on an MSA not equipped with proactive resilience mechanisms. The sudden loss of an entire node represents a systemic shock, exposing the limitations of native Kubernetes orchestration.
As shown in the upper central panel of \Cref{fig:scenario3-1}, the baseline MSA uptime fluctuates persistently between 45\% and 55\%. This indicates that the MSA spends almost half of the total time in a state of complete operational unavailability. This instability is not temporary but a permanent condition that persists for the entire experiment. This suggests that the native Kubernetes scheduler fails to stabilize the configuration after the node loss, specifically because there is no available node that can accommodate the \textit{gateway} service with its \textit{Large} flavour.

Consequently, the native orchestrator enters an infinite loop of failed scheduling attempts, repeatedly trying to reallocate orphaned services without ever converging to a valid configuration. This behaviour illustrates a fundamental limitation of standard orchestration in managing catastrophic failures.

By contrast, FREEDA's intervention demonstrates robust emergency management capabilities that overcome these structural limitations. Immediately after the perturbation, FREEDA stabilizes the MSA, restoring uptime to 100\% and maintaining it throughout the experiment (see lower central panel of \Cref{fig:scenario3-1}). Specifically, FREEDA automatically modifies the \textit{gateway} flavour and migrates it to the only suitable node, \textit{gatewaynodeefficient}, achieving both functional recovery and improved carbon efficiency.

During scenario 3.1, it was not possible to move services from the \textit{appnodestrong} node to the \textit{appnodeefficient} node, because the latter lacked sufficient resources to fully satisfy the demand.
However, FREEDA still achieved an optimal trade-off between resilience and carbon optimization. This behaviour underscores FREEDA's design principle: maximize sustainability and resource efficiency without compromising the MSA quality or availability.

\subsection{Discussion}
The analysis of the seven emulated scenarios demonstrates the effectiveness of FREEDA's multi-objective approach. In all cases, the FREEDA toolchain eliminates downtime, transforming an unstable MSA into configurations with 100\% availability. The reductions in CO2 emissions, compared to the baseline version, range from 21\% to 52\%, with an average of 35\% across all scenarios. This variability reflects the different optimization strategies adopted based on the specific characteristics of each scenario. The results highlight FREEDA's ability to manage complex scenarios, characterized by simultaneous multiple stresses and with gradual dynamic variations. In the most critical case (scenario 3.2), with a baseline uptime of 34-43\%, FREEDA achieves 100\% availability while obtaining an emission reduction of 21-24\%.

\section{Related Work}
\label{sec:related-work}

Constraint reasoning was first applied to the optimal deployment of multiservice applications on cloud resources in \cite{Zephyrus22016,DiCosmo2014,Engage2012}. Among these, \cite{Engage2012} addresses service dependency modeling, whereas \cite{Zephyrus22016} and \cite{DiCosmo2014} focus on services' hardware, software, and availability requirements. More recently, constraint reasoning has been leveraged for generating containerized MSA deployments, with \cite{Bacchiani2021} and \cite{Bravetti2019} extending the approach of \cite{Zephyrus22016} to microservice architectures, while \cite{Boreas2021} introduces container scheduling on Kubernetes based on QoS requirements. In contrast, \cite{ERASCU2021} targets the deployment of MSAs on cloud virtual machines, encoding hardware and software requirements as constraints with the objective of minimizing overall deployment costs. In summary, existing approaches typically address isolated aspects of the MSA deployment problem, such as hardware/software requirements, service dependencies, or cost optimization, and assume a complete redeployment even when contextual changes affect only part of the deployment itself. Our proposal aims to bridge these gaps by providing a holistic MSA deployment over the Cloud-Edge continuum and continuous reasoning for adaptive MSA deployment over the Cloud-Edge continuum.

Existing approaches to enforcing failure resilience in MSAs mainly provide design and development guidelines \cite{Giedrimas2018, Gremlin2016}, or mechanisms for configuring deployment scripts to self-heal failing services, typically by restarting their hosting containers \cite{brogi2022self-healing}. However, no existing solution supports the analysis of an already deployed MSA together with the available Cloud-Edge infrastructure, nor the automated enforcement of failure-resilient deployments over such infrastructures. Current techniques for failure analysis in MSAs primarily focus on detecting failures and identifying their root causes \cite{Aggarwal2021,LiuPing2020,Meng2020,Microrca2020}. While these methods can automatically infer potential root causes for an observed failure, they generally stop at this stage, leaving DevOps engineers responsible for manually inspecting logs or monitored metrics to understand how the failure propagated throughout the system \cite{Soldani2022_AnomalyDetection}. The only work moving in this direction is our previous contribution \cite{Soldani2021-WWW}, which provides DevOps engineers with explanations of failure propagation within MSAs, though it still requires them to manually specify the behavior of each service. In summary, to the best of our knowledge, no existing technique currently enables the automated analysis of an MSA deployment over a Cloud-Edge infrastructure to enforce failure resilience within that deployment. Our proposal aims to fulfill this gap by providing an explainable enhancement of MSA deployments' failure resilience.

Various existing approaches focus on improving the energy efficiency of cloud data centers \cite{GHOLIPOUR2020,VITALI2015,Wajid2016}, while best practices for enhancing data center sustainability are outlined in \cite{acton2022,singh2011data}. However, computing infrastructures are increasingly distributed across the Cloud-Edge continuum, which is inherently composed of heterogeneous nodes. This heterogeneity causes significant fluctuations in power usage effectiveness \cite{jones2018stop}, making the energy-aware allocation of Cloud-Edge resources an ongoing research challenge \cite{Xiao2017,Zhang2020}. Moreover, when applications are distributed over the Cloud-Edge continuum, their data must be stored and managed in close synergy with the applications themselves, both to reduce latency and to ensure compliance with security constraints \cite{Plebani2018}. Despite this, current approaches to improving deployment efficiency generally focus only on the target infrastructure, occasionally extending to the temporal or geographical distribution of deployed applications \cite{Nylander2018,Papadopoulos2017,Xu2021}. Other works instead emphasize the application level \cite{Alvares2010,Cappiello2011,nowak2014}, supporting the design of energy-sustainable applications from scratch, with limited applicability to existing systems. Meanwhile, the energy demand of deployed applications has grown alongside the widespread adoption of cloud computing, a trend expected to continue within Cloud-Edge infrastructures. This growing demand highlights the need for energy-aware MSA deployments across the Cloud-Edge continuum \cite{Forti2022}. Our proposal aims to fulfill this gap by providing an explainable reduction of MSA deployments' environmental impact.

\section{Conclusions}
\label{sec:conclusion}
The increasing heterogeneity, scale, and dynamism of Cloud-Edge infrastructures demand deployment strategies capable of balancing multiple, often conflicting, objectives.
In this article, we have presented the FREEDA toolchain, a comprehensive framework that automates the failure-resilient and carbon-efficient deployment of MSAs across the Cloud-Edge continuum. FREEDA integrates flavour-based optimization with adaptive orchestration strategies to dynamically reconfigure deployments in response to infrastructure variability, resource constraints, and environmental objectives.

The proposed approach enables DevOps teams to maintain service continuity and application quality while proactively minimizing carbon emissions. Through a suite of controlled simulated and emulated experiments, we have shown FREEDA's ability to mitigate failures, rebalance workloads, and adjust flavour selections to achieve an optimal compromise between resilience and sustainability. The experimental results show that the FREEDA toolchain effectively reduces environmental impact without compromising the MSA reliability or performance.

For future work, we plan to extend the support given by FREEDA by developing a full-fledged framework to support the design, development, and deployment of MSAs on the Cloud-Edge continuum. The framework will include techniques and tools for monitoring the carbon footprint of deployed MSAs, as well as to target the enhancement of other quality attributes than reliability, \eg security and performance efficiency. In this way, future releases of FREEDA will contribute to improving the environmental sustainability and overall quality of modern ICT systems.

\backmatter




\section*{Acknowledgements}
This work was supported by the \textit{FREEDA} project (CUP: I53D23003550006), funded by the frameworks PRIN (MUR, Italy) and Next Generation EU.






\bibliography{src/biblio}

\end{document}